\documentclass[twocolumn]{article}
\usepackage[T1]{fontenc}
\usepackage{kpfonts}

\usepackage{xcolor}
\definecolor{unicolor}{HTML}{990000}

\usepackage{geometry}
\usepackage{csquotes}
\usepackage[
  backend=biber,
  style=nature,
  maxcitenames=1,
  giveninits=true,
  doi=false,
  isbn=false,
  eprint=false
]{biblatex}
\let\cite\autocite
\usepackage[hidelinks]{hyperref}
\hypersetup{
    colorlinks=true,
    linkcolor=blue,
    filecolor=black,
    citecolor=unicolor,
    urlcolor=black
}

\usepackage{xurl}
\usepackage{booktabs}
\usepackage[most]{tcolorbox}
\usepackage{amssymb}
\usepackage{amsmath}

\title{The Invisible Risks of AI-Generated Health Information}

\usepackage{orcidlink}
\usepackage{authblk}
\author[1,$\dagger$]{Matthew R. DeVerna\orcidlink{0000-0003-3578-8339}}
\author[2]{Harry Yaojun Yan\orcidlink{0000-0002-9499-360X}}
\author[3]{Kai-Cheng Yang\orcidlink{0000-0003-4627-9273}}
\author[4]{Filippo Menczer\orcidlink{0000-0003-4384-2876}}

\affil[1]{Stanford University}
\affil[2]{Texas A\&M University}
\affil[3]{Binghamton University}
\affil[4]{Indiana University Bloomington}
\affil[$\dagger$]{\footnotesize To whom correspondence should be addressed: \texttt{mdeverna@stanford.edu}}

\date{
\vspace{-1em}
\small
This preprint has not yet been peer reviewed. \\
Last updated: \today
}

\newcommand{\glossterm}[1]{\textbf{\textcolor{unicolor!85!black}{#1.}}}

\begin{document}

\maketitle

\begin{abstract}
Generative artificial intelligence (AI) systems now summarize health-related search results, answer medical questions, and offer guidance people once sought from clinicians.
These systems bring real benefits, including plain-language explanations of medical information, around-the-clock availability, and expanded access for people facing language or literacy barriers.
They also carry new risks: inaccurate guidance can harm people at scale, and malicious actors can now generate personalized health misinformation at negligible cost.
In this Perspective, we argue that these risks are largely invisible to the institutions responsible for protecting public health.
When AI guidance causes harm, no record exists outside the platform, no channel allows users to report it, and no independent researcher can measure the consequences.
We trace these invisible risks across two settings: incidental exposure online and active seeking through search engines and chatbots.
Minimizing harm from AI-generated health information requires making it observable. 
We therefore offer recommendations that aim to improve transparency, mitigate harm at the point of delivery, and assign accountability, ranging from voluntary platform measures to regulatory ones.
\end{abstract}

\section*{A Shifting Landscape}
\label{sec:ecosystem}

After reading about the negative health effects of sodium chloride (table salt), a 60-year-old man consulted ChatGPT about ways to eliminate chloride from his diet.
Following advice from the AI chatbot, the man replaced sodium \textit{chloride} with sodium \textit{bromide} and was then hospitalized for paranoia and hallucinations caused by bromide toxicity~\cite{bromism}.
His physicians learned of ChatGPT only after days of treatment, when he was lucid enough to describe the substitution.
Even then, unable to retrieve the conversation itself, they concluded that they would ``never be able to know with certainty what exactly the output he received was''~\cite[(p.~2)]{bromism}. 
This case is one of a growing number of alarming reports about harmful AI-generated health information~\cite{nyt26lawsuit,nytfakehealthinfluencers}. 
Such cases are distinctive because of their invisibility: no record exists outside the AI platform---nothing for a treating clinician to correct, a regulator to act on, or a researcher to examine.

Even before the recent AI explosion, digital technologies had become deeply embedded in how people encounter health information~\cite{Haux2006HealthSysHist,Wang24082021}: social media surfaced health content without active inquiry~\cite{BorgesdoNascimento2022Infodemics,Krishnan2025Harnessing, godard2024active} and search engines returned ranked links that users could navigate and synthesize~\cite{Wallace2022Diagnostic,Hagger2021Common}.
Despite legitimate concerns about the role of platforms and algorithms in shaping health information well before generative AI~\cite{info:doi/10.2196/jmir.7.3.e33,SwireThompson2020HealthMisinfo,Obermeyer2019AIRacialBias,Pierri2022Apr}, most health information encountered online was still produced by humans, and evaluating it largely remained the responsibility of clinicians and self-directed patients.

\begin{table*}[!t]
\centering
\begin{tcolorbox}[
    enhanced,
    width=0.98\linewidth,
    colback=unicolor!3!white,
    colframe=unicolor!85!black,
    boxrule=0pt,
    frame hidden,
    borderline north={1.2pt}{0pt}{unicolor!85!black},
    borderline south={0.4pt}{0pt}{unicolor!85!black},
    sharp corners,
    left=10pt, right=10pt, top=8pt, bottom=9pt,
    title={Box 1: Glossary of key AI terms},
    fonttitle=\normalsize\bfseries,
    coltitle=white,
    colbacktitle=unicolor!85!black,
    toptitle=5pt, bottomtitle=5pt, lefttitle=10pt,
]
\small
\begin{minipage}[t]{0.48\linewidth}
\setlength{\parskip}{0.5em}\setlength{\parindent}{0pt}%
\glossterm{Generative AI} AI systems that produce novel content (text, images, audio, video) rather than only classifying or ranking existing content.

\glossterm{Large language model (LLM)} A generative AI model trained on large text corpora to produce language; underlies most current chatbots.

\glossterm{Conversational AI system (chatbot)} A user-facing application, typically built on an LLM, that engages in multi-turn dialogue. The chatbot is the product; the LLM is the model underneath.

\end{minipage}%
\hfill
\begin{minipage}[t]{0.48\linewidth}
\setlength{\parskip}{0.5em}\setlength{\parindent}{0pt}%
\glossterm{Hallucination} Fluent, confident AI output that is factually incorrect or fabricated.

\glossterm{AI-generated vs.\ AI-curated content} Content produced by generative AI versus content whose selection, ranking, or presentation to a user is determined algorithmically. Generation and curation are distinct functions, though one system may do both.

\glossterm{AI-mediated search} Search interfaces where generative AI synthesizes a direct answer instead of, or in addition to, presenting ranked lists of search results (e.g., Google's AI Overviews).

\end{minipage}
\end{tcolorbox}
\end{table*}

This arrangement is changing as generative AI, particularly large language models (LLMs), reshapes how people encounter health information (see Box~1 for a glossary of key terms).
Google search now provides AI summaries above the search results in response to health-related queries~\cite{google2023generative}, and chatbots like OpenAI's ChatGPT, Anthropic's Claude, and Google's Gemini replace those links altogether, fielding health questions at growing volume according to both independent researchers and AI platforms~\cite{AyoAjibola2024Characterizing,shen2026how}. 
OpenAI has responded to this demand with \textit{ChatGPT Health}, a ``dedicated experience in ChatGPT designed for health and wellness,'' signaling that AI-generated health information is becoming a product category of its own~\cite{openai2026gpthealth}.

Generative AI is also being integrated into clinical settings~\cite{Wachter2024WillGenAIDeliver,Shah2025Ambient}. 
Anthropic's \textit{Claude for Healthcare}~\cite{anthropic2026claudehealth} and Google's \textit{Vertex AI Search for Healthcare}~\cite{GoogleCloud2024vertexhealthsearch} serve enterprise clinical operations. 
Those deployments enter a system with existing---if imperfect---oversight structures: licensure, malpractice liability, medical-device regulation, and institutional review. 
Consumer chatbots sit outside that system entirely, despite many people asking them questions they once brought to clinicians. 
When an LLM interprets symptoms, weighs treatment options, or advises whether a condition warrants medical attention, it performs a function long treated as medical guidance.
Yet, none of the traditional structures of medical oversight exist when such guidance is generated by AI models. 
Chatbots are not the only setting in which AI creates new, invisible health risks. 
Inaccurate or harmful information can be generated and customized at scale, reaching large populations of vulnerable users through social media without provenance or attribution. 

In this Perspective, we argue that a systematic lack of visibility is a defining risk of AI-generated health information because it prevents oversight: harms are not only possible, but structurally hidden from the clinicians, researchers, and regulators who could otherwise detect and correct them. 
We develop this argument across two key settings in which the public encounters AI-generated health information outside of clinical care: incidental exposure online and active seeking through search and chatbot interfaces.
We then offer recommendations that aim to improve transparency, mitigate harm, and assign accountability.

\section*{Incidental Exposure}
\label{sec:exposure}

People encounter health information without looking for it, through routine engagement with social media and other digital technologies.
This incidental exposure is changing rapidly as generative AI is integrated into everyday information systems, with the potential for societal opportunities as well as risks.

The most immediate opportunity lies in the production of health content to which people can be exposed online. 
Generative AI expands who can produce such information, how quickly it can be created, and how precisely it can be tailored.
Institutions, advocacy groups, commercial actors, and individual creators can now generate fluent, multilingual, and audience-specific health content at minimal cost.
For public health agencies and healthcare systems, this capacity enables communication that moves beyond generic advisories toward culturally adapted and linguistically tailored messaging.
Such capabilities may improve reach among historically underserved populations and help overcome language barriers.
For instance, the World Health Organization deployed AI-powered chatbots to provide real-time guidance and counter misinformation during the COVID-19 pandemic~\cite{Panteli2025Artificial}; more capable generative systems could extend such efforts considerably in future emergencies.

Generative AI also increasingly mediates which health content people encounter and how they interpret it. 
For instance, the AI system Grok, integrated directly into the Twitter/X platform, responds to user queries about posts in public feeds~\cite{mei2026grok}. 
Those responses, including those about health, are visible to other users who never asked a question or sought out health information. 
A user scrolling past a post about vaccine safety may come away with a reinforced or revised belief, depending on how the AI system responded. 
Preliminary analysis suggests that such systems can communicate accurate guidance and surface corrections to dangerous misinformation~\cite{Renault_2025}, but other work has demonstrated LLMs' ability to launder false claims~\cite{costello2026llmconvinceconspiracytheories,DeVerna2024AIFactChecking}.
As AI-generated and AI-curated health content proliferates into more digital spaces, it will play an increasingly consequential role in determining which health claims people accept as true.

AI systems can shape information exposure through curation. 
Researchers and developers have begun to explore whether LLM-based systems can rerank social media feeds to surface more accurate and pro-social content, given their capacity to interpret meaning and context~\cite{Piccardi_2025}. 
Such tools could take the form of browser extensions that sit on top of commercial platforms, or conversational interfaces built directly into those platforms, allowing users to actively shape what they encounter~\cite{stray2026prosocialranking}.
For health information, the implications cut both ways: a user who instructs such a system to prioritize high-quality sources could construct an information environment substantially more accurate than what commercial algorithms provide, while one nudged toward untrustworthy content could accelerate the adoption of fringe health beliefs, as suggested by preliminary experiments~\cite{costello2026llmconvinceconspiracytheories}. 

Even exposure to health content generated with the best intention can be problematic, because AI systems personalize and curate health content faster than anyone can review it.
Public health agencies that once vetted a small set of advisories before release cannot feasibly check the millions of individualized variants AI systems might produce for different audiences, languages, and contexts.
Feed-integrated AI systems pose the same problem, because no one can audit responses generated on demand at global scale.
Whether personalized or generated on demand, AI-generated health information can propagate subtle inaccuracies, outdated guidance, and miscalibrated recommendations as error at scale.

At the same time, generative AI dramatically lowers the cost and effort for malicious actors to expose people to harmful health content.
Such actors may include commercial operators promoting unproven remedies and state-sponsored actors with strategic interests in weakening population health, both of whom can flood digital spaces with targeted messaging at minimal cost.
Some social media accounts already consist largely or entirely of AI-generated text, images, or video, giving rise to ``AI influencers'' whose authoritative-sounding but synthetic health messages reach millions of users~\cite{Saeidnia_2026}, as widely reported in the news~\cite{guardian2025deepfakes,Myers2025,Silverman2026, nytfakehealthinfluencers}. 
Coordinated groups of autonomous agents, or AI ``swarms,'' can manufacture the appearance of many seemingly independent voices converging on the same health claim~\cite{Schroeder2026swarms}. 
These AI tools give state actors a new means of harming target populations, for example, by persuading them to forego vaccination during a pandemic~\cite{DeVerna2024ModelMisinfoDisease}. 

\section*{Active Seeking}
\label{sec:seeking}

When people start to actively seek out health information, they will find that Google and other search engines now present AI summaries above ranked results, synthesizing information from the web to directly answer their queries~\cite{Wardle2025Evolving}. 
Chatbot systems such as ChatGPT, Claude, and Gemini go even further, replacing ranked links altogether with conversational interfaces~\cite{FernandezPichel2025Evaluating}.
Rather than retrieving documents for users to read and evaluate, these systems interpret and summarize them, providing users with easy-to-digest, synthesized responses.
Conversational AI tools are thus replacing traditional search engines as the first point of contact when people seek health information~\cite{Hopkins2023AIChatbots}. 
About one-third of U.S. adults report using AI interfaces for health information or advice as of March 2026~\cite{Montero2026KFF}.
News accounts also report that some people are using chatbots to self-diagnose~\cite{NYTWell2026,nyt26lawsuit}.
Like the opening case study, these accounts illustrate how AI systems blur the boundary between information summary and care-oriented judgment, potentially shaping symptom interpretation and downstream decision-making~\cite{FernandezPichel2025Evaluating,Hopkins2023AIChatbots}.

AI-aided symptom interpretation creates an opportunity to solve a real problem. 
Authoritative health content online is frequently written for clinicians rather than patients, and physicians tend to rely on jargon that leaves patients confused or reluctant to ask follow-up questions~\cite{daraz2018readability,gotlieb2022accuracy}.
AI systems can translate this complex medical information into plain language calibrated to the user's level of understanding.
They can also synthesize across a far broader range of sources than a person would typically consult, compressing what might otherwise be hours of reading into a single response.
Even for patients who leave clinical encounters without fully understanding their diagnosis or treatment options, these systems supply what a static web page cannot: an interlocutor that explains, re-explains, and adjusts its language on demand~\cite{romoff2025readability}.
Chatbots may also be the only affordable source of medical advice available to people facing language barriers, limited health literacy, or no access to health care~\cite{Montero2026KFF}.

Conversational interfaces also remove friction and social pressure that affects whether people ask health questions at all.
Users can raise questions at any hour, without waiting or embarrassment.  
And unlike a hurried appointment at the doctor's office or ER visit~\cite{hoek2020discharge}, a well-functioning AI system can deliver responses for users to consume at their preferred pace. 
In a nationally representative survey of U.S. adults conducted in early 2026, 65\% of those who used AI for health advice cited wanting quick or immediate information as a major reason, and 36\% cited feeling more comfortable looking up health questions privately~\cite{Montero2026KFF}. 

These strengths come with corresponding fragility.
AI systems can produce fluent, confident statements that are factually wrong~\cite{huang2025hallucination}.
Because coherence is a property of the language model rather than a guarantee of accuracy, users may mistake polished language for reliability and forgo verification when it matters most~\cite{Hopkins2023AIChatbots}.
In health contexts, such errors can distort symptom interpretation or perceived urgency~\cite{Ramaswamy_2026}.
The same push toward synthesis can also compress nuance, collapsing qualified, context-dependent guidance into simplified claims that are easier to read but more likely to mislead.

AI systems also reflect knowledge cutoffs based on when they are trained~\cite{cacioli2025do}.
While much health guidance is relatively stable, recommendations do change; a system unaware of updated clinical guidance will present outdated information with the same confidence it applies to settled knowledge.
Even when AI systems use retrieval or web search to mitigate this risk, the capacity to draw on many sources simultaneously does not guarantee that reliable medical evidence will be prioritized~\cite{DeVerna2025LLMBenchmarking,omar2026mapping,hu2026auditinggooglesaioverviews}, and models may fail to recognize unreliable sources at all~\cite{yang2025accuracy}.
The resulting synthesis can therefore obscure disagreement or evidence gaps~\cite{FernandezPichel2025Evaluating}.
Worse, this weakness can be exploited deliberately. 
Adversaries can publish false claims on websites that are then unwittingly absorbed by models during training, a practice known as ``data poisoning''~\cite{alber2025medical,abtahi2026datapoisoning,walker2026scientists}. 

Privacy is another concern. 
AI platforms have an unprecedented ability to collect sensitive health data from user conversations, such as symptoms, diagnoses, and personal circumstances. 
This information may be retained, used for model training, monetized through targeted advertising, sold to third parties, or leaked in breaches. 
Such uses create new risks such as insurance discrimination and identity theft. 
Health information enjoys high levels of protection in traditional medical settings to protect against such risks~\cite{10.1001/jama.2023.9458}, but mainstream AI chatbots are currently outside such regulations.

These risks are compounded by inconsistency.
Performance varies across AI providers, models, and even prompt phrasing, so the quality of AI-generated health guidance is neither uniform nor predictable~\cite{FernandezPichel2024BinaryQA}.
Whether a user receives sound guidance depends in part on which system they happen to consult and how they happen to phrase the question.

\section*{Invisible Harms}
\label{sec:harms}

Repeated encounters with fluent, personalized health messages during routine digital activity can shape background beliefs about risk, prevention, and treatment~\cite{udry2024illusory}.
When users act on the information they receive, the downstream consequences can extend to delayed care, inappropriate reassurance or escalation~\cite{Ramaswamy_2026}.
For example, experimental evidence demonstrates that anti-vaccine messaging can increase vaccine hesitancy~\cite{Loomba_2021}. 
Such content measurably suppressed vaccination uptake and contributed to increased disease burden during COVID-19~\cite{Pierri2022Apr,Bollenbacher_2026,DeVerna2024ModelMisinfoDisease}.
These risks can be weaponized through generative AI, changing the scale at which malicious users can produce such content and the precision with which it can be tailored.
Large language models can also be exploited to durably influence attitudes~\cite{Costello2024DurablyReduce}, including on medical topics such as HPV vaccination, according to preliminary findings~\cite{Xu2025HPV}.

The distribution of these harms may exacerbate existing digital divides. 
People facing language barriers, limited health literacy, or no reliable access to care are among the most exposed to AI health guidance~\cite{Montero2026KFF} and among the least well served by it. 
For example, non-English responses are substantially less accurate and less consistent than their English counterparts~\cite{Jin2024BetterEnglish}. 
Deploying imperfect systems to underserved communities may also blunt incentives to expand human-delivered care, widening the very disparities AI promises to reduce.

When established forms of health guidance cause harm, they leave records that can be traced: a clinician who advises a patient produces a chart entry; a harmful drug generates an adverse-event report; a misleading advertisement remains publicly visible for regulators to act upon.
The risks of AI cataloged above---unpredictable performance, outdated guidance, hallucinations, poisoned models, and malicious exploitation---rarely leave any of these records.
This lack of systematic visibility impedes attribution.

Accountability requires attributing a health outcome to a specific AI output, but two obstacles make that attribution nearly impossible.
First, an exchange with a chatbot is visible only to the company that operates the model, and AI companies provide no API access for collecting and studying such data.
For self-deployed open-source models, conversations are only visible to individual users.
Second, fluent and confident prose gives users no signal that anything went wrong, so errors that would otherwise surface as complaints stay silent.
A user who is wrongly reassured that their symptoms are benign, for example, may never connect a later delayed diagnosis to the exchange that produced it.
Such harms surface even less among vulnerable populations who are less studied, less present in follow-up care, and have fewer resources for recourse.

The circulation of AI-generated health claims on social media compounds these problems of provenance and attribution.
A chatbot user at least knows they received AI-generated guidance; someone scrolling past an AI-generated health claim may never learn what produced it.
And researchers cannot measure the exposure, since platforms have withdrawn the access that once permitted study of how content spreads~\cite{freelon2024post,ortutay2024crowdtangle}.
Ironically, AI models themselves contributed to this lack of access by making user-generated text valuable as training data~\cite{Mimizuka2025PostPostAPI}.

\section*{Governance Recommendations}
\label{sec:discussion}

\begin{table*}[t]
    \centering
    \caption{Governance gaps in AI-generated health information, organized by objective, with corresponding remedies and the stakeholders positioned to deliver them.}
    \small
    \renewcommand{\arraystretch}{1.25}
    \begin{tabular}{@{}p{0.13\linewidth}p{0.16\linewidth}p{0.46\linewidth}p{0.14\linewidth}@{}}
        \toprule
        \textbf{Goal} & \textbf{Gap} & \textbf{Recommendation} & \textbf{Stakeholder} \\
        \midrule
        \textbf{Transparency} &
        Lack of records &
        Access rights to one's own health interactions with AI systems, so that any individual harm can be shared with clinicians and reconstructed after the fact &
        AI platforms \\
        \addlinespace
        &
        Lack of disclosure &
        Systematic disclosure allowing AI health guidance to be reported, aggregated, and investigated, including concerning events flagged internally (analogous to adverse-event reporting) &
        AI platforms \\
        \addlinespace
        &
        No independent evaluation &
        Data access allowing independent, ongoing evaluation of deployed systems, including across languages and populations &
        AI and social-media platforms \\
        \midrule
        \textbf{Mitigation} &
        Lack of provenance &
        Watermarking, AI detection, and clear labeling of AI-generated health content; withholding such content from wide distribution until experts have vetted it and its provenance is established &
        Social-media platforms \\
        \addlinespace
        &
        Inaccurate health guidance &
        Treatment of health queries as a distinct, higher-stakes category; AI responses and summaries drawn only from sources vetted by medical governing bodies &
        AI and search-engine platforms \\
        \addlinespace
        &
        Lack of e-health literacy accounting for generative-AI risks &
        Caution toward AI-generated health content rather than deference to it; consultation of original sources rather than reliance on AI summaries &
        Users \\
        \midrule
        \textbf{Accountability} &
        Lack of liability for harms caused by AI-generated health information &
        Extension of physician malpractice liability to chatbots used for medical guidance; in the U.S., removal of Section~230 immunity for harmful AI-generated health content &
        Policymakers \\
        \bottomrule
    \end{tabular}
    \label{tab:gaps}
\end{table*}

A system cannot be governed while the harms it produces remain unobservable.
We therefore treat observability as a necessary precondition for meaningful oversight of AI-generated health information~\cite{Pan_2026}.
Our recommendations proceed accordingly, from transparency to mitigation to accountability.
Table~\ref{tab:gaps} outlines the gaps we have identified, potential remedies, and the actors in a position to deliver them.

Existing regulations build transparency into traditional health care systems.
When a patient wants to see what a clinician wrote, the law gives them a right to it~\cite{ESSEN201844}.
When a drug harms someone, an adverse-event report enters a public record, and reports accumulate until a pattern emerges that no individual clinician could have seen~\cite{BENINGER20181991,DangyCaye2025pharmaregulation}.
No equivalent rules cover AI-generated health content.
Users should have a right of access to their own health interactions with an AI system, so that they can reconstruct a harmful exchange and show it to a clinician.
Those exchanges should then be reported, aggregated, and investigated the same way adverse medical events are, so that harms recorded one at a time can be understood systematically.
AI providers should contribute to that record as well, systematically disclosing how their models handle health queries and which exchanges they flag internally.
Every major AI company has the technical means to build these systems today.

Disclosure alone will not be enough.
A provider that reports on itself decides what counts as a harm worth reporting, and we should not expect commercial firms to report problematic cases against their own interest.
External benchmarks avoid that conflict, but they measure a system before anyone uses it and tell us little about how these tools perform at scale in the hands of real users~\cite{Angus2025AI}.
Independent researchers therefore need direct and ongoing access to deployed AI systems, and that access must cover all of the languages and populations those systems serve.
The European Union already imposes such access requirements on very large online platforms, but compliance has fallen well short of what the law requires~\cite{goanta2026great,Mimizuka2025PostPostAPI}.
Access rights on paper are not access in practice, and any regulation should anticipate that gap.
Transparency measures also widen access to sensitive health data, and must carefully specify who gains access, to what, and under what protections.

Mitigating harm calls for measures beyond transparency.
Unreliable AI-generated health content now reaches people who never asked for it, and nothing about it reveals its provenance.
Rules governing the promotion of prescription pharmaceuticals confronted a similar problem before generative AI.
Drug manufacturers cannot advertise directly to U.S. consumers without identifying the message as promotional and stating risks alongside benefits~\cite{CFRPrescriptionAds, Alves2019}.
The European Union's AI Act applies similar logic to synthetic content: providers of generative AI systems must mark their outputs in machine-readable form, and platforms must label deepfakes as well as AI-generated text published on matters of public interest~\cite{AIAct2024}.
For AI-generated medical content, governments worldwide should impose equivalent disclosure requirements.
Platforms should not wait for legislation.
We recommend that they detect, watermark, and label synthetic health content, and withhold it from wide distribution until experts have vetted it and established its provenance.

AI platforms also control how their models answer, and a change to that behavior takes effect across every query at once.
At present, a question about chest pain is handled like a question about restaurant hours.
We recommend that platforms treat health queries as a distinct, higher-stakes category and answer them only from sources vetted by professional medical societies.

Even with such measures in place, users should treat AI-generated health information with caution and consult original sources.
Such literacy efforts, however, place the burden of evaluation on the general public, who are poorly equipped to assess medical content and vulnerable to a fluent but inaccurate answer.
Reliability must be built into AI-generated health content before it reaches the public.

Reliability does not remove the need for accountability.
When a clinician's advice harms a patient, malpractice law determines who answers for it; no equivalent covers a chatbot.
In the United States, Section~230 immunizes platforms against liability for user-generated content.
Because a user's prompt shapes what a chatbot says, whether that immunity extends to AI summaries and chatbot answers remains unsettled.
Courts elsewhere have begun to weigh in.
In May 2026, the Regional Court of Munich ordered Google to stop repeating false claims its AI Overviews had made about specific publishers~\cite{transparencycoalition2026munich}.
The court reasoned that AI Overviews evaluate, combine, and rewrite existing material into new statements rather than pointing users toward it.
Treating those statements as Google's own, the court placed them outside the protections that cover traditional search, and it rejected the argument that advising users to verify results removes the provider's responsibility for what those results say.
Although the ruling is preliminary, contested, and concerned with reputational rather than medical harm, its reasoning applies to anything that rewrites sources into new statements, which is what a chatbot does every time it answers a health question.
We recommend that U.S. legislators reach the same conclusion by statute rather than leaving it to litigation, and that physician malpractice liability be extended to chatbots offering medical guidance.

Legislation of this kind may take years~\cite{Ong_2024,Mesko2023}, whereas platforms could adopt most of the measures we propose today.
Until they do, these harms remain invisible by choice rather than necessity.

\section*{Acknowledgments}

The authors thank Richard Street Jr. for discussions that informed this perspective.
We also acknowledge the use of generative AI (Gemini and Claude Code) to run preliminary literature searches and copyedit the manuscript; the authors wrote all text, verified all cited sources, and take full responsibility for this work.

\section*{Funding}

F.M. discloses partial support from the Knight Foundation.
M.R.D., H.Y.Y, and K-C.Y. declare no relevant funding.

\section*{Competing Interests}

The authors declare no competing interests. 

\section*{Data and Code Availability}

No original data or code were generated for this article.

\printbibliography

@article{ESSEN201844,
title = {Patient access to electronic health records: Differences across ten countries},
author = {Anna Essén and Isabella Scandurra and Reinie Gerrits and Gayl Humphrey and Monika Alise Johansen and Patrick Kierkegaard and Jani Koskinen and Siaw-Teng Liaw and Souad Odeh and Peeter Ross and Jessica S. Ancker},
journal = {Health Policy and Technology},
volume = {7},
number = {1},
pages = {44--56},
year = {2018},
doi = {10.1016/j.hlpt.2017.11.003},
url = {https://doi.org/10.1016/j.hlpt.2017.11.003},
}

@article{BENINGER20181991,
    title = {Pharmacovigilance: An Overview},
    author = {Paul Beninger},
    journal = {Clinical Therapeutics},
    volume = {40},
    number = {12},
    pages = {1991--2004},
    year = {2018},
    doi = {10.1016/j.clinthera.2018.07.012},
    url = {https://doi.org/10.1016/j.clinthera.2018.07.012},
}

@ARTICLE{DangyCaye2025pharmaregulation,    
    AUTHOR={Dangy-Caye, Agnes  and Mousset, Alice  and Kermad, Adem  and Bouché-Bazerolle, Louise  and Bujar, Magda  and De Lucia, Maria-Lucia  and McAuslane, Neil  and Lumsden, Rebecca },
    TITLE={Harmonizing health: a global analysis of pharmaceutical regulatory activities by international regulatory organizations},
    JOURNAL={Frontiers in Medicine}, 
    VOLUME={12},
    YEAR={2025},  
    DOI={10.3389/fmed.2025.1636269},
    url = {https://doi.org/10.3389/fmed.2025.1636269},
}

@article{Alves2019,
	author = {Leonardo Alves, Teresa and Lexchin, Joel and Mintzes, Barbara},
	title = {Medicines Information and the Regulation of the Promotion of Pharmaceuticals},
	journal = {Science and Engineering Ethics},
	volume = {25},
	number = {4},
	pages = {1167--1192},
    doi = {10.1007/s11948-018-0041-5},
    url = {https://doi.org/10.1007/s11948-018-0041-5},
	year = {2019}
}

@misc{CFRPrescriptionAds,
  author = {{U.S. Food and Drug Administration}},
  title = {{Prescription-Drug Advertisements}},
  howpublished = {21 C.F.R. \S~202.1},
  year = {2026},
  note = {Accessed 2026-08-16},
  url = {https://www.ecfr.gov/current/title-21/chapter-I/subchapter-C/part-202/section-202.1}
}

@misc{AIAct2024,
  author = {{European Parliament and Council of the European Union}},
  title = {{Regulation (EU) 2024/1689 (Artificial Intelligence Act)}},
  howpublished = {OJ L, 2024/1689, 12.7.2024},
  year = {2024},
  note = {CELEX: 32024R1689. Accessed 2026-08-16},
  url = {http://data.europa.eu/eli/reg/2024/1689/oj}
}

@article{info:doi/10.2196/jmir.7.3.e33,
    author={Walther, Joseph B and Pingree, Suzanne
    and Hawkins, Robert P and Buller, David B},
    title={Attributes of Interactive Online Health Information Systems},
    journal={J Med Internet Res},
    year={2005},
    volume={7},
    number={3},
    pages={e33},
    doi={10.2196/jmir.7.3.e33},
    url={https://doi.org/10.2196/jmir.7.3.e33}
}

@article{godard2024active,
  title={Are active and passive social media use related to mental health, wellbeing, and social support outcomes? A meta-analysis of 141 studies},
  author={Godard, Rebecca and Holtzman, Susan},
  journal={Journal of Computer-Mediated Communication},
  volume={29},
  number={1},
  pages={zmad055},
  year={2024},
  publisher={Oxford University Press},
  doi = {https://doi.org/10.1093/jcmc/zmad055},
  url = {https://doi.org/10.1093/jcmc/zmad055},
}

@article{Wang24082021,
    author = {Xiaohui Wang and Jingyuan Shi and Hanxiao Kong},
    title = {Online Health Information Seeking: A Review and Meta-Analysis},
    journal = {Health Communication},
    volume = {36},
    number = {10},
    pages = {1163--1175},
    year = {2021},
    publisher = {Routledge},
    doi = {10.1080/10410236.2020.1748829},
    note ={PMID: 32290679},
    URL = {https://doi.org/10.1080/10410236.2020.1748829},
    eprint = {https://doi.org/10.1080/10410236.2020.1748829}
}

@misc{nyt26lawsuit,
  author       = {Teddy Rosenbluth},
  title        = {ChatGPT Led to a Man’s Near-Fatal Health Crisis, Lawsuit Claims},
  howpublished = {The New York Times},
  year         = {2026},
  month        = jul,
  day          = {22},
  url          = {https://www.nytimes.com/2026/07/22/well/openai-chatgpt-health-lawsuit.html},
  note         = {Accessed: 2026-07-24}
}

@article{10.1001/jama.2023.9458,
    author = {Marks, Mason and Haupt, Claudia E.},
    title = {{AI Chatbots, Health Privacy, and Challenges to HIPAA Compliance}},
    journal = {JAMA},
    volume = {330},
    number = {4},
    pages = {309--310},
    year = {2023},
    doi = {10.1001/jama.2023.9458},
    url = {https://doi.org/10.1001/jama.2023.9458}
}

@article{bromism,
    author = {Eichenberger, Audrey and Thielke, Stephen and Van Buskirk, Adam}, 
    title = {A Case of Bromism Influenced by Use of Artificial Intelligence},
    journal = {Annals of Internal Medicine: Clinical Cases},
    volume = {4},
    number = {8},
    pages = {e241260},
    year = {2025},
    doi = {10.7326/aimcc.2024.1260},
    URL = {https://doi.org/10.7326/aimcc.2024.1260},
}

@article{abtahi2026datapoisoning,
  author   = {Abtahi, Farhad
              and Seoane, Fernando
              and Pau, Ivan
              and Vega-Barbas, Mario},
  title    = {Data Poisoning Vulnerabilities Across Health Care Artificial Intelligence Architectures: Analytical Security Framework and Defense Strategies},
  journal  = {Journal of Medical Internet Research},
  year     = {2026},
  month    = {1},
  day      = {23},
  volume   = {28},
  pages    = {e87969},
  doi      = {10.2196/87969},
  url = {https://doi.org/10.2196/87969},
}

@article{alber2025medical,
  author    = {Alber, Daniel Alexander and Yang, Zihao and Alyakin, Anton and Yang, Eunice and Rai, Sumedha and Valliani, Aly A. and Zhang, Jeff and Rosenbaum, Gabriel R. and Amend-Thomas, Ashley K. and Kurland, David B. and Kremer, Caroline M. and Eremiev, Alexander and Negash, Bruck and Wiggan, Daniel D. and Nakatsuka, Michelle A. and Sangwon, Karl L. and Neifert, Sean N. and Khan, Hammad A. and Save, Akshay Vinod and Palla, Adhith and Grin, Eric A. and Hedman, Monika and Nasir-Moin, Mustafa and Liu, Xujin Chris and Jiang, Lavender Yao and Mankowski, Michal A. and Segev, Dorry L. and Aphinyanaphongs, Yindalon and Riina, Howard A. and Golfinos, John G. and Orringer, Daniel A. and Kondziolka, Douglas and Oermann, Eric Karl},
  doi       = {10.1038/s41591-024-03445-1},
  journal   = {Nature Medicine},
  month     = jan,
  number    = {2},
  pages     = {618--626},
  publisher = {Springer Science and Business Media LLC},
  title     = {Medical large language models are vulnerable to data-poisoning attacks},
  url       = {http://dx.doi.org/10.1038/s41591-024-03445-1},
  volume    = {31},
  year      = {2025}
}

@article{Angus2025AI,
  author = {Angus, Derek C. and Khera, Rohan and Lieu, Tracy and Liu, Vincent and Ahmad, Faraz S. and Anderson, Brian and Bhavani, Sivasubramanium V. and Bindman, Andrew and Brennan, Troyen and Celi, Leo Anthony and Chen, Frederick and Cohen, I. Glenn and Denniston, Alastair and Desai, Sanjay and Embí, Peter and Faisal, Aldo and Ferryman, Kadija and Gerhart, Jackie and Gross, Marielle and Hernandez-Boussard, Tina and Howell, Michael and Johnson, Kevin and Lee, Kristine and Liu, Xiaoxuan and Lomis, Kimberly and London, Alex John and Longhurst, Christopher A. and Mandl, Kenneth D. and McGlynn, Elizabeth and Mello, Michelle M. and Munoz, Fatima and Ohno-Machado, Lucila and Ouyang, David and Perlis, Roy and Phillips, Adam and Rhew, David and Ross, Joseph S. and Saria, Suchi and Schwamm, Lee and Seymour, Christopher W. and Shah, Nigam H. and Shah, Rashmee and Singh, Karandeep and Solomon, Matthew and Spates, Kathryn and Spector-Bagdady, Kayte and Wang, Tommy and Gichoya, Judy Wawira and Weinstein, James and Wiens, Jenna and Bibbins-Domingo, Kirsten and Alterovitz, Gil and Clancy, Heather A and Dawson, Lindsay and Diamond, Matthew and Holve, Erin C and Kahn, Jeremy and Pengetnze, Yolande M and Rao, Shiv and Shrank, William H and Termulo, Cesar},
  journal = {JAMA},
  month = Nov,
  number = {18},
  pages = {1650},
  publisher = {American Medical Association (AMA)},
  title = {{AI, Health, and Health Care Today and Tomorrow: The JAMA Summit Report on Artificial Intelligence}},
  url = {http://dx.doi.org/10.1001/jama.2025.18490},
  volume = {334},
  year = {2025}
}

@misc{anthropic2026claudehealth,
  author       = {Anthropic},
  title        = {{Advancing Claude in healthcare and the life sciences}},
  year         = {2026},
  month        = apr,
  howpublished = {Public Press Release},
  note         = {Accessed: 2026-04-17},
  url          = {https://www.anthropic.com/news/healthcare-life-sciences}
}

@misc{shen2026how,
  author       = {Shen, Judy Hanwen and Carter, Shan and Dargan, Richard and Gillotte, Jessica and Handa, Kunal and Hong, Jerry and Huang, Saffron and Jagadish, Kamya and Kearney, Matt and Levinstein, Ben and Linthicum, Ryn and McCain, Miles and Millar, Thomas and Julapalli, Mo and Price, Sara and Stern, Michael and Saunders, David and Tamkin, Alex and Vallone, Andrea and Clark, Jack and Pollack, Sarah and Eaton, Jake and Ganguli, Deep and Durmus, Esin},
  title        = {{How people ask Claude for personal guidance}},
  year         = {2026},
  howpublished = {Anthropic Research},
  url          = {https://www.anthropic.com/research/claude-personal-guidance},
  note         = {Accessed: 2026-05-07}
}

@article{AyoAjibola2024Characterizing,
  title   = {{Characterizing the Adoption and Experiences of Users of Artificial Intelligence-Generated Health Information in the United States: Cross-Sectional Questionnaire Study}},
  volume  = {26},
  url     = {https://doi.org/10.2196/55138},
  doi     = {10.2196/55138},
  journal = {Journal of Medical Internet Research},
  author  = {Ayo-Ajibola, Oluwatobiloba and Davis, Ryan J. and Lin, Matthew E. and Riddell, Jeffrey and Kravitz, Richard L.},
  year    = {2024},
  pages   = {e55138}
}

@article{Bollenbacher_2026,
  author    = {Bollenbacher, John and Menczer, Filippo and Bryden, John},
  journal   = {EPJ Data Science},
  month     = jan,
  number    = {1},
  publisher = {Springer Science and Business Media LLC},
  title     = {{Effects of antivaccine tweets on COVID-19 vaccinations, cases, and deaths}},
  url       = {http://dx.doi.org/10.1140/epjds/s13688-025-00606-1},
  volume    = {15},
  year      = {2026}
}

@article{BorgesdoNascimento2022Infodemics,
  title   = {{Infodemics and health misinformation: a systematic review of reviews}},
  volume  = {100},
  url     = {https://doi.org/10.2471/BLT.21.287654},
  doi     = {10.2471/BLT.21.287654},
  number  = {9},
  journal = {Bulletin of the World Health Organization},
  author  = {Borges do Nascimento, Israel Júnior and Pizarro, Ana Beatriz and Almeida, Jussara M. and Azzopardi-Muscat, Natasha and Gonçalves, Marcos André and Björklund, Maria and Novillo-Ortiz, David},
  month   = jun,
  year    = {2022},
  pages   = {544--561}
}

@inproceedings{cacioli2025do,
  title     = {Do Knowledge Cutoffs Drive Clinical Accuracy? Quantifying Temporal Decay in Large Language Models},
  author    = {Michael Cacioli and Aryan Arya and Austen Liao and Kevin Zhu},
  booktitle = {Socially Responsible and Trustworthy Foundation Models at NeurIPS 2025},
  year      = {2025},
  url       = {https://openreview.net/forum?id=z5p3MlzAs5}
}

@article{Costello2024DurablyReduce,
  author    = {Costello, Thomas H. and Pennycook, Gordon and Rand, David G.},
  journal   = {Science},
  month     = sep,
  number    = {6714},
  publisher = {American Association for the Advancement of Science (AAAS)},
  title     = {{Durably reducing conspiracy beliefs through dialogues with AI}},
  url       = {http://dx.doi.org/10.1126/science.adq1814},
  volume    = {385},
  year      = {2024}
}

@article{daraz2018readability,
  author  = {Lubna Daraz and Allison S. Morrow and Oscar J. Ponce and Wigdan Farah and Abdulrahman Katabi and Abdul Majzoub and Mohamed O. Seisa and Raed Benkhadra and Mouaz Alsawas and Prokop Larry and M. Hassan Murad},
  title   = {Readability of Online Health Information: A Meta-Narrative Systematic Review},
  journal = {American Journal of Medical Quality},
  volume  = {33},
  number  = {5},
  pages   = {487-492},
  year    = {2018},
  doi     = {10.1177/1062860617751639},
  url = {https://doi.org/10.1177/1062860617751639},
  note    = {PMID: 29345143}
}

@article{DeVerna2024ModelMisinfoDisease,
  author  = {DeVerna, Matthew R. and Pierri, Francesco and Ahn, Yong-Yeol and Fortunato, Santo and Flammini, Alessandro and Menczer, Filippo},
  title   = {Modeling the amplification of epidemic spread by individuals exposed to misinformation on social media},
  journal = {npj Complexity},
  volume  = {2},
  number  = {11},
  pages   = {1--8},
  year    = {2025},
  url     = {https://doi.org/10.1038/s44260-025-00038-y}
}

@misc{DeVerna2025LLMBenchmarking,
  title = {Large Language Models Require Curated Context for Reliable Political Fact-Checking -- Even with Reasoning and Web Search},
  author = {Matthew R. DeVerna and Kai-Cheng Yang and Harry Yaojun Yan and Filippo Menczer},
  year = {2025},
  eprint = {2511.18749},
  archiveprefix = {arXiv},
  primaryclass  = {cs.CL},
  url = {https://arxiv.org/abs/2511.18749}
}

@inproceedings{FernandezPichel2024BinaryQA,
  address   = {Cham},
  title     = {{Large Language Models for Binary Health-Related Question Answering: A Zero- and Few-Shot Evaluation}},
  doi       = {10.1007/978-3-031-63772-8_29},
  url = {https://doi.org/10.1007/978-3-031-63772-8_29},
  booktitle = {Computational {Science} -- {ICCS} 2024},
  publisher = {Springer Nature Switzerland},
  author    = {Fernández-Pichel, Marcos and Losada, David E. and Pichel, Juan C.},
  editor    = {Franco, Leonardo and de Mulatier, Clélia and Paszynski, Maciej and Krzhizhanovskaya, Valeria V. and Dongarra, Jack J. and Sloot, Peter M. A.},
  year      = {2024},
  pages     = {325--339}
}

@article{FernandezPichel2025Evaluating,
  title   = {Evaluating search engines and large language models for answering health questions},
  volume  = {8},
  url     = {https://www.nature.com/articles/s41746-025-01546-w},
  doi     = {10.1038/s41746-025-01546-w},
  number  = {1},
  journal = {npj Digital Medicine},
  author  = {Fernández-Pichel, Marcos and Pichel, Juan C. and Losada, David E.},
  month   = mar,
  year    = {2025},
  pages   = {153}
}

@misc{google2023generative,
  author       = {{Elizabeth Reid}},
  title        = {{Supercharging Search with Generative AI}},
  year         = {2023},
  month        = may,
  howpublished = {{The Keyword | Google Official Blog}},
  note         = {Accessed: 2025-05-23},
  url          = {https://blog.google/products/search/generative-ai-search/}
}

@misc{nytfakehealthinfluencers,
  author = {Arijeta Lajka and Isabelle Niu and Mark Boyer and James Surdam and Dan T. Peters},
  title = {{The Fake Influencers Selling Wellness on Your Feed}},
  howpublished = {The New York Times},
  year = {2026},
  month = jul,
  url = {https://www.nytimes.com/video/technology/100000011001849/ai-influencers-health-supplements-fake-ads.html},
  note = {Accessed: 2026-08-06}
}

@article{gotlieb2022accuracy,
  author  = {Gotlieb, Rachael and Praska, Corinne and Hendrickson, Marissa A. and Marmet, Jordan and Charpentier, Victoria and Hause, Emily and Allen, Katherine A. and Lunos, Scott and Pitt, Michael B.},
  title   = {Accuracy in Patient Understanding of Common Medical Phrases},
  journal = {JAMA Network Open},
  volume  = {5},
  number  = {11},
  pages   = {e2242972-e2242972},
  year    = {2022},
  month   = {11},
  issn    = {2574-3805},
  doi     = {10.1001/jamanetworkopen.2022.42972},
  url = {https://doi.org/10.1001/jamanetworkopen.2022.42972},
}

@misc{guardian2025deepfakes,
  author       = {Denis Campbel},
  title        = {AI deepfakes of real doctors spreading health misinformation on social media},
  howpublished = {The Guardian},
  year         = {2025},
  month        = dec,
  day          = {5},
  url          = {https://www.theguardian.com/society/2025/dec/05/ai-deepfakes-of-real-doctors-spreading-health-misinformation-on-social-media},
  note         = {Accessed: 2026-03-27}
}

@article{Hagger2021Common,
  title   = {{The common sense model of illness self-regulation: a conceptual review and proposed extended model}},
  volume  = {15},
  url     = {https://doi.org/10.1080/17437199.2021.1878050},
  doi     = {10.1080/17437199.2021.1878050},
  number  = {4},
  journal = {Health Psychology Review},
  author  = {Hagger, Martin S. and Orbell, Sheina},
  year    = {2021},
  pages   = {494--538}
}

@article{Haux2006HealthSysHist,
  title     = {{Health information systems --- past, present, future}},
  volume    = {75},
  url       = {http://dx.doi.org/10.1016/j.ijmedinf.2005.08.002},
  doi       = {10.1016/j.ijmedinf.2005.08.002},
  number    = {3--4},
  journal   = {International Journal of Medical Informatics},
  publisher = {Elsevier BV},
  author    = {Haux, Reinhold},
  year      = {2006},
  month     = mar,
  pages     = {268--281}
}

@article{hoek2020discharge,
  title   = {Patient Discharge Instructions in the Emergency Department and Their Effects on Comprehension and Recall of Discharge Instructions: A Systematic Review and Meta-analysis},
  journal = {Annals of Emergency Medicine},
  author  = {Amber E. Hoek and Susanne C.P. Anker and Ed F. {van Beeck} and Alex Burdorf and Pleunie P.M. Rood and Juanita A. Haagsma},
  volume  = {75},
  number  = {3},
  pages   = {435-444},
  year    = {2020},
  doi     = {https://doi.org/10.1016/j.annemergmed.2019.06.008},
  url = {https://doi.org/10.1016/j.annemergmed.2019.06.008},
}

@article{Hopkins2023AIChatbots,
  title   = {{Artificial intelligence chatbots will revolutionize how cancer patients access information: ChatGPT represents a paradigm-shift}},
  volume  = {7},
  url     = {https://doi.org/10.1093/jncics/pkad010},
  doi     = {10.1093/jncics/pkad010},
  number  = {2},
  urldate = {2025-12-07},
  journal = {JNCI Cancer Spectrum},
  author  = {Hopkins, Ashley M and Logan, Jessica M and Kichenadasse, Ganessan and Sorich, Michael J},
  month   = apr,
  year    = {2023},
  pages   = {pkad010}
}

@inproceedings{Jin2024BetterEnglish,
  address   = {New York, NY, USA},
  series    = {{WWW '24}},
  title     = {{Better to Ask in English: Cross-Lingual Evaluation of Large Language Models for Healthcare Queries}},
  url       = {https://dl.acm.org/doi/10.1145/3589334.3645643},
  doi       = {10.1145/3589334.3645643},
  urldate   = {2025-09-21},
  booktitle = {Proceedings of the {ACM} {Web} {Conference} 2024},
  publisher = {Association for Computing Machinery},
  author    = {Jin, Yiqiao and Chandra, Mohit and Verma, Gaurav and Hu, Yibo and De Choudhury, Munmun and Kumar, Srijan},
  month     = may,
  year      = {2024},
  pages     = {2627--2638}
}

@inproceedings{Krishnan2025Harnessing,
  title     = {{Harnessing AI for Innovative Marketing Strategies in the Healthcare Sector}},
  booktitle = {2025 IEEE International Conference on Data Science and Information System (ICDSIS)},
  publisher = {IEEE},
  author    = {Krishnan, M. Murali and Manoharan, Geetha and Selvaraj, Franklin John and Shiva, Roshan Kumar},
  year      = {2025}
}

@article{Loomba_2021,
  author    = {Loomba, Sahil and de Figueiredo, Alexandre and Piatek, Simon J. and de Graaf, Kristen and Larson, Heidi J.},
  journal   = {Nature Human Behaviour},
  month     = feb,
  number    = {3},
  pages     = {337--348},
  publisher = {Springer Science and Business Media LLC},
  title     = {{Measuring the impact of COVID-19 vaccine misinformation on vaccination intent in the UK and USA}},
  url       = {http://dx.doi.org/10.1038/s41562-021-01056-1},
  volume    = {5},
  year      = {2021}
}

@misc{mei2026grok,
  title         = {Grok in the Wild: Characterizing the Roles and Uses of Large Language Models on Social Media},
  author        = {Katelyn Xiaoying Mei and Robert Wolfe and Nicholas Weber and Martin Saveski},
  year          = {2026},
  eprint        = {2602.11286},
  archiveprefix = {arXiv},
  primaryclass  = {cs.SI},
  url           = {https://arxiv.org/abs/2602.11286}
}

@article{Mesko2023,
  author    = {Meskó, Bertalan and Topol, Eric J.},
  journal   = {npj Digital Medicine},
  month     = jul,
  number    = {1},
  publisher = {Springer Science and Business Media LLC},
  title     = {{The imperative for regulatory oversight of large language models (or generative AI) in healthcare}},
  url       = {http://dx.doi.org/10.1038/s41746-023-00873-0},
  volume    = {6},
  year      = {2023}
}

@misc{Montero2026KFF,
  author       = {Montero, Alex and Montalvo III, Julian and Kearney, Audrey and Valdes, Isabelle and Kirzinger, Ashley and Hamel, Liz},
  title        = {{KFF Tracking Poll on Health Information and Trust: Use of AI For Health Information and Advice}},
  howpublished = {KFF Poll},
  day          = {25},
  month        = mar,
  year         = {2026},
  url          = {https://www.kff.org/public-opinion/kff-tracking-poll-on-health-information-and-trust-use-of-ai-for-health-information-and-advice/},
  note         = {Accessed: 2026-04-02}
}

@misc{NYTWell2026,
  author       = {Maggie Astor},
  title        = {{Patients Turn to AI for Medical Answers When Doctors Fall Short}},
  howpublished = {The New York Times},
  year         = {2026},
  month        = apr,
  day          = {2},
  url          = {https://www.nytimes.com/2026/04/02/well/live/ai-illness-claude-chatgpt.html},
  note         = {Accessed: 2026-04-06}
}

@article{Obermeyer2019AIRacialBias,
  author    = {Obermeyer, Ziad and Powers, Brian and Vogeli, Christine and Mullainathan, Sendhil},
  doi       = {10.1126/science.aax2342},
  journal   = {Science},
  month     = oct,
  number    = {6464},
  pages     = {447--453},
  publisher = {American Association for the Advancement of Science (AAAS)},
  title     = {{Dissecting racial bias in an algorithm used to manage the health of populations}},
  url       = {http://dx.doi.org/10.1126/science.aax2342},
  volume    = {366},
  year      = {2019}
}

@article{omar2026mapping,
  author    = {Omar, Mahmud and Sorin, Vera and Wieler, Lothar H and Charney, Alexander W and Kovatch, Patricia and Horowitz, Carol R and Korfiatis, Panagiotis and Glicksberg, Benjamin S and Freeman, Robert and Nadkarni, Girish N and Klang, Eyal},
  doi       = {10.1016/j.landig.2025.100949},
  issn      = {2589-7500},
  journal   = {The Lancet Digital Health},
  month     = jan,
  number    = {1},
  pages     = {100949},
  publisher = {Elsevier BV},
  title     = {Mapping the susceptibility of large language models to medical misinformation across clinical notes and social media: a cross-sectional benchmarking analysis},
  url       = {http://dx.doi.org/10.1016/j.landig.2025.100949},
  volume    = {8},
  year      = {2026}
}

@misc{openai2026gpthealth,
  author       = {OpenAI},
  title        = {{Introducing ChatGPT Health}},
  year         = {2026},
  month        = jan,
  howpublished = {Public Press Release},
  note         = {Accessed: 2026-01-31},
  url          = {https://openai.com/index/introducing-chatgpt-health/}
}

@article{Panteli2025Artificial,
  title   = {{Artificial intelligence in public health: promises, challenges, and an agenda for policy makers and public health institutions}},
  volume  = {10},
  url     = {https://doi.org/10.1016/S2468-2667(25)00036-2},
  journal = {The Lancet Public Health},
  author  = {Panteli, Dimitra and Adib, Keyrellous and Buttigieg, Stefan and Goiana-da-Silva, Francisco and Ladewig, Katharina and Azzopardi-Muscat, Natasha and Figueras, Josep and Novillo-Ortiz, David and McKee, Martin},
  month   = feb,
  year    = {2025},
  pages   = {e428--e432}
}

@article{Piccardi_2025,
  author    = {Piccardi, Tiziano and Saveski, Martin and Jia, Chenyan and Hancock, Jeffrey and Tsai, Jeanne L. and Bernstein, Michael S.},
  journal   = {Science},
  month     = nov,
  number    = {6776},
  publisher = {American Association for the Advancement of Science (AAAS)},
  title     = {{Reranking partisan animosity in algorithmic social media feeds alters affective polarization}},
  url       = {http://dx.doi.org/10.1126/science.adu5584},
  volume    = {390},
  year      = {2025}
}

@article{Pierri2022Apr,
  author  = {Pierri, Francesco and Perry, Brea L. and DeVerna, Matthew R. and Yang, Kai-Cheng and Flammini, Alessandro and Menczer, Filippo and Bryden, John},
  title   = {{Online misinformation is linked to early COVID-19 vaccination hesitancy and refusal}},
  journal = {Sci Rep},
  volume  = {12},
  number  = {5966},
  pages   = {1--7},
  year    = {2022},
  month   = apr,
  doi     = {10.1038/s41598-022-10070-w},
  url     = {https://doi.org/10.1038/s41598-022-10070-w}
}

@article{Ramaswamy_2026,
  author    = {Ramaswamy, Ashwin and Tyagi, Alvira and Hugo, Hannah and Jiang, Joy and Jayaraman, Pushkala and Jangda, Mateen and Te, Alexis E. and Kaplan, Steven A. and Lampert, Joshua and Freeman, Robert and Gavin, Nicholas and Tewari, Ashutosh K. and Sakhuja, Ankit and Naved, Bilal and Charney, Alexander W. and Omar, Mahmud and Gorin, Michael A. and Klang, Eyal and Nadkarni, Girish N.},
  journal   = {Nature Medicine},
  month     = feb,
  publisher = {Springer Science and Business Media LLC},
  title     = {{ChatGPT Health performance in a structured test of triage recommendations}},
  url       = {http://dx.doi.org/10.1038/s41591-026-04297-7},
  year      = {2026}
}

@misc{Renault_2025,
  author       = {Renault, Thomas and Mosleh, Mohsen and Rand, David Gertler},
  month        = dec,
  howpublished = {Center for Open Science},
  title        = {{@Grok Is This True? LLM-Powered Fact-Checking on Social Media}},
  url          = {http://dx.doi.org/10.31234/osf.io/85quw_v1},
  year         = {2025}
}

@article{romoff2025readability,
  author  = {Romoff, Melissa and Brunette, Madison and Peterson, Melanie K. and Hashmi, Sohaib Z. and Kim, Michael S.},
  title   = {The role of large language models in improving the readability of orthopaedic spine patient educational material},
  journal = {Journal of Orthopaedic Surgery and Research},
  year    = {2025},
  volume  = {20},
  number  = {1},
  pages   = {531},
  doi     = {10.1186/s13018-025-05955-1},
  url = {https://doi.org/10.1186/s13018-025-05955-1},
  month   = may,
  day     = {28}
}

@article{Schroeder2026swarms,
  author    = {Schroeder, Daniel Thilo and Cha, Meeyoung and Baronchelli, Andrea and Bostrom, Nick and Christakis, Nicholas A. and Garcia, David and Goldenberg, Amit and Kyrychenko, Yara and Leyton-Brown, Kevin and Lutz, Nina and Marcus, Gary and Menczer, Filippo and Pennycook, Gordon and Rand, David G. and Ressa, Maria and Schweitzer, Frank and Song, Dawn and Summerfield, Christopher and Tang, Audrey and Van Bavel, Jay J. and van der Linden, Sander and Kunst, Jonas R.},
  journal   = {Science},
  month     = jan,
  number    = {6783},
  pages     = {354--357},
  publisher = {American Association for the Advancement of Science (AAAS)},
  title     = {{How malicious AI swarms can threaten democracy}},
  url       = {http://dx.doi.org/10.1126/science.adz1697},
  volume    = {391},
  year      = {2026}
}

@article{Shah2025Ambient,
  title   = {{Ambient artificial intelligence scribes: physician burnout and perspectives on usability and documentation burden}},
  volume  = {32},
  url     = {https://doi.org/10.1093/jamia/ocae295},
  doi     = {10.1093/jamia/ocae295},
  number  = {2},
  journal = {Journal of the American Medical Informatics Association},
  author  = {Shah, Shreya J. and Devon-Sand, Anna and Ma, Stephen P. and Jeong, Yejin and Crowell, Trevor and Smith, Margaret and Liang, April S. and Delahaie, Clarissa and Hsia, Caroline and Shanafelt, Tait and Pfeffer, Michael A. and Sharp, Christopher and Lin, Steven and Garcia, Patricia},
  month   = dec,
  year    = {2025},
  pages   = {375--380}
}

@misc{stray2026prosocialranking,
  archiveprefix = {arXiv},
  title         = {The Prosocial Ranking Challenge: Reducing Polarization on Social Media without Sacrificing Engagement},
  author        = {Jonathan Stray and Ian Baker and George Beknazar-Yuzbashev and Ceren Budak and Julia Kamin and Kylan Rutherford and Mateusz Stalinski and Tin Acosta and Chris Bail and Michael Bernstein and Mark Brandt and Amy Bruckman and Anshuman Chhabra and Soham De and Kayla Duskin and Sara Fish and Beth Goldberg and Andy Guess and Dylan Hadfield-Menell and Muhammed Haroon and Safwan Hossain and Michael Inzlicht and Gauri Jain and Yanchen Jiang and Alexander P. Landry and Yph Lelkes and Hongfan Lu and Peter Mason and Jennifer McCoy and Smitha Milli and Paul Resnick and Emily Saltz and Martin Saveski and Lisa Schirch and Max Spohn and Siddarth Srinivasan and Alexis Tatore and Luke Thorburn and Joshua A. Tucker and Robb Willer and Magdalena Wojcieszak and Manuel Wüthrich and Sylvan Zheng},
  year          = {2026},
  eprint        = {2603.19626},
  archiveprefix = {arXiv},
  primaryclass  = {cs.SI},
  url           = {https://arxiv.org/abs/2603.19626}
}

@article{SwireThompson2020HealthMisinfo,
  author    = {Swire-Thompson, Briony and Lazer, David},
  doi       = {10.1146/annurev-publhealth-040119-094127},
  journal   = {Annual Review of Public Health},
  month     = apr,
  number    = {1},
  pages     = {433--451},
  publisher = {Annual Reviews},
  title     = {{Public Health and Online Misinformation: Challenges and Recommendations}},
  url       = {http://dx.doi.org/10.1146/annurev-publhealth-040119-094127},
  volume    = {41},
  year      = {2020}
}

@article{udry2024illusory,
  title   = {The illusory truth effect: A review of how repetition increases belief in misinformation},
  journal = {Current Opinion in Psychology},
  volume  = {56},
  pages   = {101736},
  year    = {2024},
  issn    = {2352-250X},
  doi     = {https://doi.org/10.1016/j.copsyc.2023.101736},
  url     = {https://www.sciencedirect.com/science/article/pii/S2352250X23001811},
  author  = {Jessica Udry and Sarah J. Barber}
}

@article{Wachter2024WillGenAIDeliver,
  author    = {Wachter, Robert M. and Brynjolfsson, Erik},
  doi       = {10.1001/jama.2023.25054},
  journal   = {JAMA},
  month     = jan,
  number    = {1},
  pages     = {65},
  publisher = {American Medical Association (AMA)},
  title     = {{Will Generative Artificial Intelligence Deliver on Its Promise in Health Care?}},
  url       = {http://dx.doi.org/10.1001/jama.2023.25054},
  volume    = {331},
  year      = {2024}
}

@article{walker2026scientists,
  author    = {Stokel-Walker, Chris},
  doi       = {10.1038/d41586-026-01100-y},
  journal   = {Nature},
  month     = apr,
  publisher = {Springer Science and Business Media LLC},
  title     = {Scientists invented a fake disease. AI told people it was real},
  url       = {http://dx.doi.org/10.1038/d41586-026-01100-y},
  year      = {2026}
}

@article{Wallace2022Diagnostic,
  title   = {{The diagnostic and triage accuracy of digital and online symptom checker tools: a systematic review}},
  volume  = {5},
  url     = {https://doi.org/10.1038/s41746-022-00667-w},
  doi     = {10.1038/s41746-022-00667-w},
  journal = {npj Digital Medicine},
  author  = {Wallace, William and Chan, Calvin and Chidambaram, Swathikan and Hanna, Lydia and Iqbal, Fahad Mujtaba and Acharya, Amish and Normahani, Pasha and Ashrafian, Hutan and Markar, Sheraz R. and Sounderajah, Viknesh and Darzi, Ara},
  year    = {2022},
  pages   = {118}
}

@article{Wardle2025Evolving,
  title   = {{Evolving Health Information-Seeking Behavior in the Context of Google AI Overviews, ChatGPT, and Alexa: Interview Study Using the Think-Aloud Protocol}},
  volume  = {27},
  url     = {https://www.jmir.org/2025/1/e79961},
  doi     = {10.2196/79961},
  number  = {1},
  urldate = {2025-10-14},
  journal = {Journal of Medical Internet Research},
  author  = {Wardle, Claire and Urbani, Shaydanay and Wang, Eric},
  month   = oct,
  year    = {2025},
  pages   = {e79961}
}

@misc{Xu2025HPV,
  author = {Xu, Henry George and Costello, Thomas H and Schwartz, Jason L. and Niccolai, Linda M. and Pennycook, Gordon and Rand, David Gertler},
  howpublished = {Preprint[OSF]},
  title = {{Personalized AI Dialogues and Parents' HPV Vaccine Intent: A Randomized Trial}},
  url = {http://dx.doi.org/10.31234/osf.io/gv52j_v1},
  year = {2025}
}

@article{Pan_2026,
  author = {Pan, Jiazhen and Hager, Paul and Schlager, Moritz and Thompson, Humphrey and McCradden, Melissa and Rueckert, Daniel},
  journal = {Nature Health},
  month = feb,
  number = {2},
  pages = {162--163},
  title = {{COMPASS-GH is a consensus roadmap for defining standards for safe, accurate and equitable AI in general health queries}},
  url = {http://dx.doi.org/10.1038/s44360-026-00053-w},
  volume = {1},
  year = {2026}
}

@misc{Myers2025,
  author = {Myers, Steven Lee and Callahan, Alice and Rosenbluth, Teddy},
  title = {{The Doctors Are Real, but the Sales Pitches Are Frauds}},
  howpublished = {The New York Times},
  year = {2025},
  month = sep,
  day = {5},
  url = {https://www.nytimes.com/2025/09/05/technology/ai-doctor-scams.html},
  note = {Accessed: 2026-04-23}
}

\end{document}